\documentclass[epj]{svjour}

\usepackage{amsmath,amssymb,psfig,epsfig,wrapfig} 

\usepackage[english]{babel}
\newcommand{\eqeqref}[1]{Eq.~(\ref{#1})}
\newcommand{\eqseqref}[1]{Eqs.~\eqref{#1}}
\newcommand{\refref}[1]{Ref.~\cite{#1}}

\newcommand{\figref}[1]{Fig.~\ref{#1}}

\newcommand{\etal}{{\it et al.}}

\newcommand{\beq}{\begin{equation}}
\newcommand{\eeq}{\end{equation}}
\newcommand{\bea}{\begin{eqnarray}}
\newcommand{\beas}{\begin{eqnarray*}}
\newcommand{\beau}[1]{\begin{equation} \label{#1} \begin{array}{rcl}}
\newcommand{\eea}{\end{eqnarray}}
\newcommand{\eeas}{\end{eqnarray*}}
\newcommand{\eeau}{\end{array} \end{equation}}
\newcommand{\bay}{\begin{array}}
\newcommand{\eay}{\end{array}}
\newcommand{\bal}{\begin{align}}
\newcommand{\eal}{\end{align}}
\newcommand{\bals}{\begin{align*}}
\newcommand{\eals}{\end{align*}}

\newcommand{\ra}{{\rightarrow}}

\newcommand{\vev}[1]{\langle #1 \rangle}

\newcommand{\In}[1]{ \bigg| _{#1} }

\begin{document}
\title{``Naked'' Cronin effect in A+A collisions from SPS to RHIC}
\author{
  Alberto Accardi
  \inst{1}
  \thanks{Talk given at ``Hard Probes 2004'', Ericeira (Portugal), November 4-10, 2004}
} 
\institute{Dept. of Physics and Astronomy, Iowa State U., Ames, IA
  50011, USA}
\date{}
\abstract{
Baseline computations of the Cronin effect in nuclear collisions
at energies spanning the SPS and the RHIC accelerators 
are performed in the Glauber-Eikonal model, which ascribes the
effect to initial-state incoherent multiple parton scatterings. The model 
accounts very well for the mid-rapidity Cronin effect in
hadron-nucleus collisions in the $\sqrt{s}=27-200$ GeV center of mass
energy range, and will be extended to nucleus-nucleus collisions. 
The computations are performed under the assumption that the partons
do not interact  with the medium produced in the collision.
Therefore, medium effects
such as energy loss in a Quark-Gluon Plasma may be detected and measured
as deviations from the presented baseline computation of the ``naked''
Cronin effect. 
} 
\maketitle
\section{Introduction}
\label{sec:intro}

The Cronin effect  \cite{Cronin} 
is the deformation of hadron transverse momentum
spectrum in hadron-nucleus (p+A) and nucleus-nucleus (A+A) collisions 
relative to linear extrapolation from proton-proton (p+p) reactions.
The Cronin effect may be quantified by taking the hadron transverse momentum
($p_T$) spectrum in p+A collision in a given centrality class (c.cl.),
normalizing it to binary scaled p+p collisions by the inverse thickness
function $T_A$, and finally dividing it by the p+p spectrum:
\begin{equation}
  R_{hA} =  
    {\frac{1}{T_A(c.cl.)} \frac{dN}{dp_T^2 dy}
       ^{\hspace{-0.3cm}pA\rightarrow h X}
       \hspace{-0.8cm} (c.cl.)} \Bigg/
    {\frac{d\sigma}{dp_T^2 dy}
       ^{\hspace{-0.3cm}pp\rightarrow h X}
       \hspace{-0.8cm} } \ .
  \label{CroninRatio}
\end{equation}
For A+A collisions, an analogous definition of $R_{AA}$ is used. 
To cancel systematic errors as much as possible, it is also customary
to take the ratio of a given centrality class to the most
peripheral one. 

In p+A collisions, the ratio $R_{hA}$ is smaller than 1 at low
momentum $p_T$$\lesssim$2 GeV (the spectrum is suppressed). 
At intermediate $p_T$$\sim$2-6 GeV is
larger than one (the spectrum is enhanced) and finally tends to 1 at
higher $p_T$'s (the spectrum follows a p+p scaling).
The described bell shape of the ``Cronin ratio'' $R_{hA}$ 
has been observed on a wide energy range $\sqrt s \approx 20-200$ and
for different hadron species \cite{Cronin,CroninRHIC}.

In A+A at low energy $\sqrt{s}=17.4$ GeV at SPS \cite{CroninAA_SPS}, 
one observes a similar behaviour of the Cronin ratio as in p+A 
collisions. However, at RHIC energy
$\sqrt{s}=63-200$ GeV, the above described bell shape appears on top 
of a largely suppressed spectrum \cite{WA98-PHENIX}. This suppression 
is often interpreted as due to energy loss of a parton traveling a
medium of density up to 15 times the normal nuclear density
\cite{enloss}, largely sufficient for the creation of a Quark-Gluon
Plasma (QGP). 

An intriguing fact in both the p+A and A+A cases is the much larger
magnitude of 
the effect on baryons compared to mesons, with the
difference between the two seeming to decrease with energy
\cite{Cronin,bmanomaly}. 
This aspect of the famed ``baryon/meson'' anomaly in high
energy nuclear collision has not
yet found an accepted explanation in theory, and will later come back
into our discussion.

In this note, I discuss the Glauber-Eikonal (GE) model computation
\cite{AG04} of the Cronin effect in A+A collisions. 
The model describes multiple parton scatterings in pQCD and
extrapolates the physics of 
proton-proton (p+p) collisions to p+A and A+A collisions without ad
hoc parameters for the nuclear case. This allows a baseline
computation of the Cronin effect in A+A without the further
complications of medium induced modifications of the $p_T$ spectrum.
This is what I call the bare, or ``naked'', Cronin effect.
Comparison with experimental data allows to measure the medium induced
effects, and in particular to detect when the medium begins to
suppress hadron spectra as a function of the energy and centrality of the
collisions. 

\section{The Glauber-Eikonal model}
\label{sec:GE}

The Glauber-Eikonal (GE) approach \cite{AG04} to the Cronin
effect treats multiple $2\ra 2$ partonic collisions in collinearly
factorized pQCD. The low-$p_T$ spectra in nuclear collisions are
suppressed by unitarity. At moderate $p_T$, the accumulation  of
transverse momentum  leads to an enhancement of transverse spectra. At
high $p_T$ the binary scaled p+p spectrum is recovered.

The cross-section for the production of a hadron $h$ with
transverse momentum $p_T$ and rapidity $y$ in $p+A$ collisions at fixed
impact parameter $b$ is written as 
\begin{align}
  \frac{d\sigma}{d^2p_T dy d^2b}^{\hspace{-0.5cm}pA\rightarrow hX}
    \hspace{-0.5cm} & =
    \vev{xf_{i/p}} \otimes \frac{d\sigma^{\,iA}}{d^2p_T dy_i d^2b} 
    \otimes D_{i\ra h} \, \In{y_h = y}  
 \label{pAxsec} \\ \nonumber
  & + T_A(b) \vev{xf_{j/A}} \otimes \frac{d\sigma^{\,jp}}{d^2p_T dy_j}  
    \otimes D_{j\ra h} \, \In{y_h = -y} 
\end{align}
where the crossed-circle symbols denote appropriate integrations and
summations over parton flavours $i$ and $j$, see \refref{AG04} for details..
The first term in \eqeqref{pAxsec} accounts for multiple semihard
scatterings of the parton 
$i$ on the target nucleus; in  the second term the nucleus
partons are assumed to undergo a single scattering on the proton.
$T_A(b)$ is the target nucleus thickness function, and $D_{i\ra h}$
is the fragmentation function of the $i$ parton.
The average parton flux from the proton, $\vev{xf_{i/p}}$,  
and the parton-nucleon cross-section, $d\sigma^{\,iN}$,
are defined as
\begin{align}
  \frac{d\sigma^{iN}}{d^2p_T dy_i} & = 
    K\, \frac{d\hat\sigma}{d\hat t}^{ij}\!\!(p_0)
    \otimes \, x_jf_{j/N} (\vev{k_T^2}) 
\label{iNxsec} \\
  \vev{xf_{i/p}} & = K\, x_i f_{i/p}(\vev{k_T^2}) 
    \otimes \frac{d\hat\sigma}{d\hat t}^{ij}\!\!(p_0)
    \otimes \, x_j f_{j/N}(\vev{k_T^2}) \nonumber \\ 
  & \times \left( \frac{d\sigma^{iN}}{d^2p_T dy_i} \right)^{-1}
 \label{avfluxandxsec} 
\end{align}
where only the sum over $j$ is understood. 
$\hat t$ is the Mandelstam variable and  $d\hat\sigma/d\hat t$ are
leading order parton-parton cross-sections in collinearly factorized
pQCD. $K$ is a constant factor which takes into account next-to-leading
order corrections. 
To regularize the IR divergences of the single-scattering pQCD parton-nucleon
cross-sections, a small mass regulator $p_0$ is introduced in the
propagators, and $Q=\sqrt{p_T^2+p_0^2}/2$ is the scale of the hard process. 
Finally, a small intrinsic transverse
momentum $\vev{k_T^2}=0.52$ GeV$^2$ is introduced to better describe the hadron
spectra in p+p collisions at intermediate $p_T \approx 2-5$ GeV. 
Since all parameters have been fixed in p+p collisions, we are able to 
compute the spectra in p+A and A+A collision with no extra freedom.

The free parameters $p_0$ and $K$ in Eqs.~\eqref{iNxsec}-\eqref{avfluxandxsec})
are fitted to, viz., low- and high-$p_T$ hadron production data in
p+p collisions at the energy and rapidity of interest. For this
study, I took from \refref{AG04} the values of $p_0$ extracted at 
$\sqrt s = 27.4$ and $200$ GeV, viz., $p_0=0.7$ and $1.0$ GeV, 
and assumed a logarithmic dependence on the energy:
\begin{align}
  p_0 = 0.151 + 0.100 \log\left(\frac{s}{\small 1\ \text{GeV}}\right) 
    \,\pm\, 10\% \ \text{GeV} \, .
 \label{p0extrap}
\end{align} 
As regards the $K$ factor, it turns out to be $K\approx 1$ quite
insensitive to the energy of the collision in the 27-200 GeV energy
range, and I will use this value throughout this note. 
Clearly, one would need more systematic studies of spectra in p+p
collisions, especially when extrapolating the parameters at the SPS energy
of 17 GeV. These will be presented elsewhere.

Nuclear effects are included in $d\sigma^{\,iA}$, 
the average transverse momentum distribution of a 
parton who suffered at least one semihard scattering on the target
nucleus A:
\begin{align}
  \frac{d\sigma^{\,iA}}{d^2p_Tdyd^2b}
    & = \sum_{n=1}^{\infty} \frac{1}{n!} \int d^2b \, d^2k_1 \cdots d^2k_n
    \, \delta\big(\sum _{i=1,n} {\vec k}_i - {\vec p_T}\big) 
    \nonumber \\
  & \times \frac{d\sigma^{\,iN}}{d^2k_1} T_A(b) 
    \times \dots \times \frac{d\sigma^{\,iN}}{d^2k_n} T_A(b) 
    \nonumber \\
  & \times e^{\, - \sigma^{\,iN}(p_0) T_A(b)} \, 
 \label{dWdp}
\end{align}
This equation sums all processes with $n$ multiple $2\ra 2$
parton scatterings. The exponential factor in \eqeqref{dWdp}
represents the probability that the parton suffered no semihard
scatterings after the $n$-th one, and explicitly unitarizes the
cross-section at the nuclear level. 

\begin{figure}
\psfig{figure=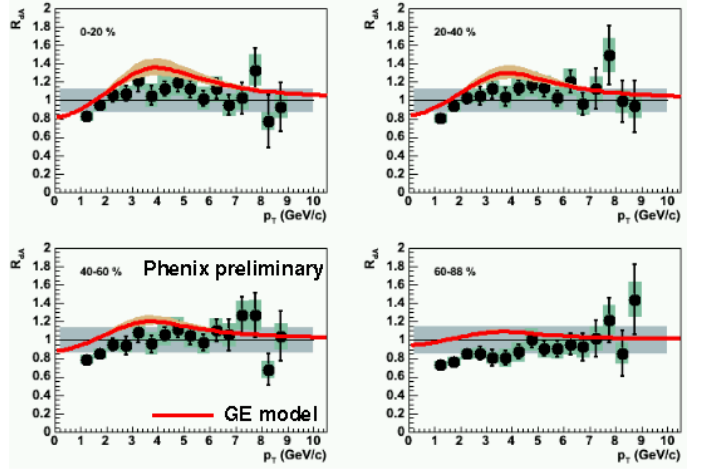,width=0.5\textwidth}
\caption{Cronin effect on neutral pion production 
as a function of centrality in d+Au collisions
at RHIC. Curved lines and band are the GE model computation with its
theoretical uncertainty. The horizontal gray band is the experimental
uncertainty in the normalization of the ratio. 
Figure adapted from \refref{Bueschingtalk}.
}
\label{fig:PHENIX_dAu}
\end{figure}

The above presented GE model describes quite well the energy and
centrality dependence of the Cronin effect on pion production 
in p+A collisions at Fermilab $\sqrt{s}=27.4$ GeV, and in d+Au
collisions at RHIC $\sqrt{s}=200$ GeV, see \refref{AG04} and
\figref{fig:PHENIX_dAu}. As all other models based on multiple
initial-state parton scatterings, the model fails in describing the
effect on baryon production. The main reason is that baryon production
is not understood in this framework even at the p+p level, which is
the starting point of the GE approach, see \eqseqref{iNxsec} and
\eqref{avfluxandxsec}. An alternative explanation of the
baryon/meson anomaly based on final-state parton recombination
\cite{Hwa} is more successful in this respect than the independent
fragmentation approach used in the GE model.

\section{The ``naked'' Cronin effect in A+A collisions}
\label{sec:naked}

The natural generalization of \eqeqref{pAxsec} to the case of
nucleus-nucleus collisions is to allow the partons from the target
nucleus to rescatter on the projectile nucleus, as well:
\begin{align}
  \frac{d\sigma}{d^2p_T dy d^2b}^{\hspace{-0.5cm}AB\rightarrow hX}
    \hspace{-0.5cm} & = \int d^2s\, \label{AAxsec} \\
  & \hspace*{-1.34cm} \times \Bigg\{ T_A(b-s) \vev{xf_{i/A}} 
    \otimes \frac{d\sigma^{\,iB}}{d^2p_T dy_i d^2s} 
    \otimes D_{i\ra h} \, \In{y_h = y}  
 \nonumber \\ 
  & \hspace*{-1.4cm} + T_B(s) \vev{xf_{j/B}} \otimes \frac{d\sigma^{\,iA}}{d^2p_T dy_i
  d^2(b-s)} \otimes D_{j\ra h} \, \In{y_h = -y} \Bigg\}
 \nonumber
\end{align}
As no medium effects are included in this formula, what we are
computing is the ``naked'' Cronin effect, i.e., the effect
stripped of medium-induced modifications such as, e.g., jet quenching.
Hence, the resulting Cronin ratio may be used as a baseline to detect
and measure jet quenching. 

\begin{figure}[t]
\epsfig{figure=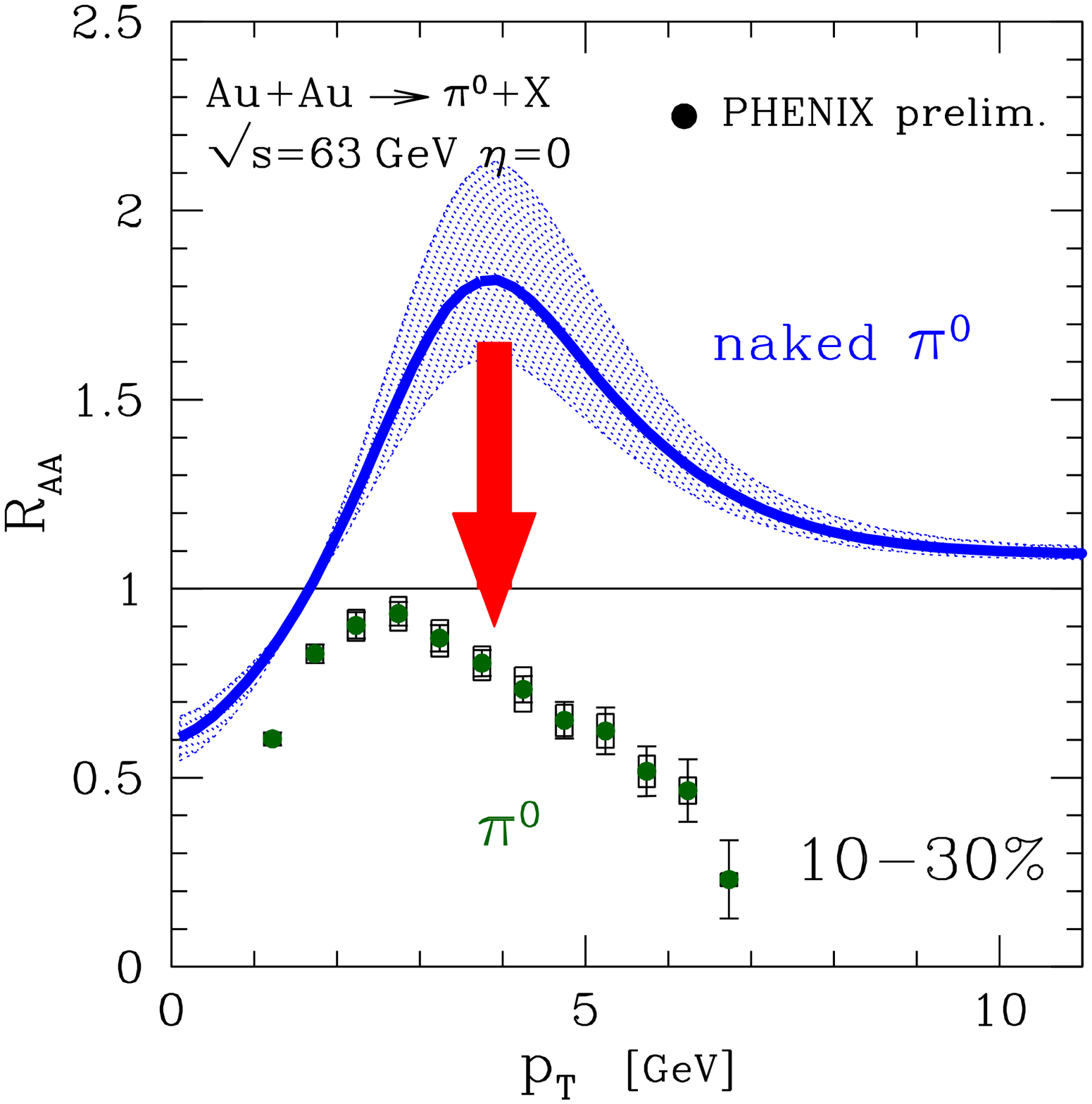,width=0.23\textwidth}
\epsfig{figure=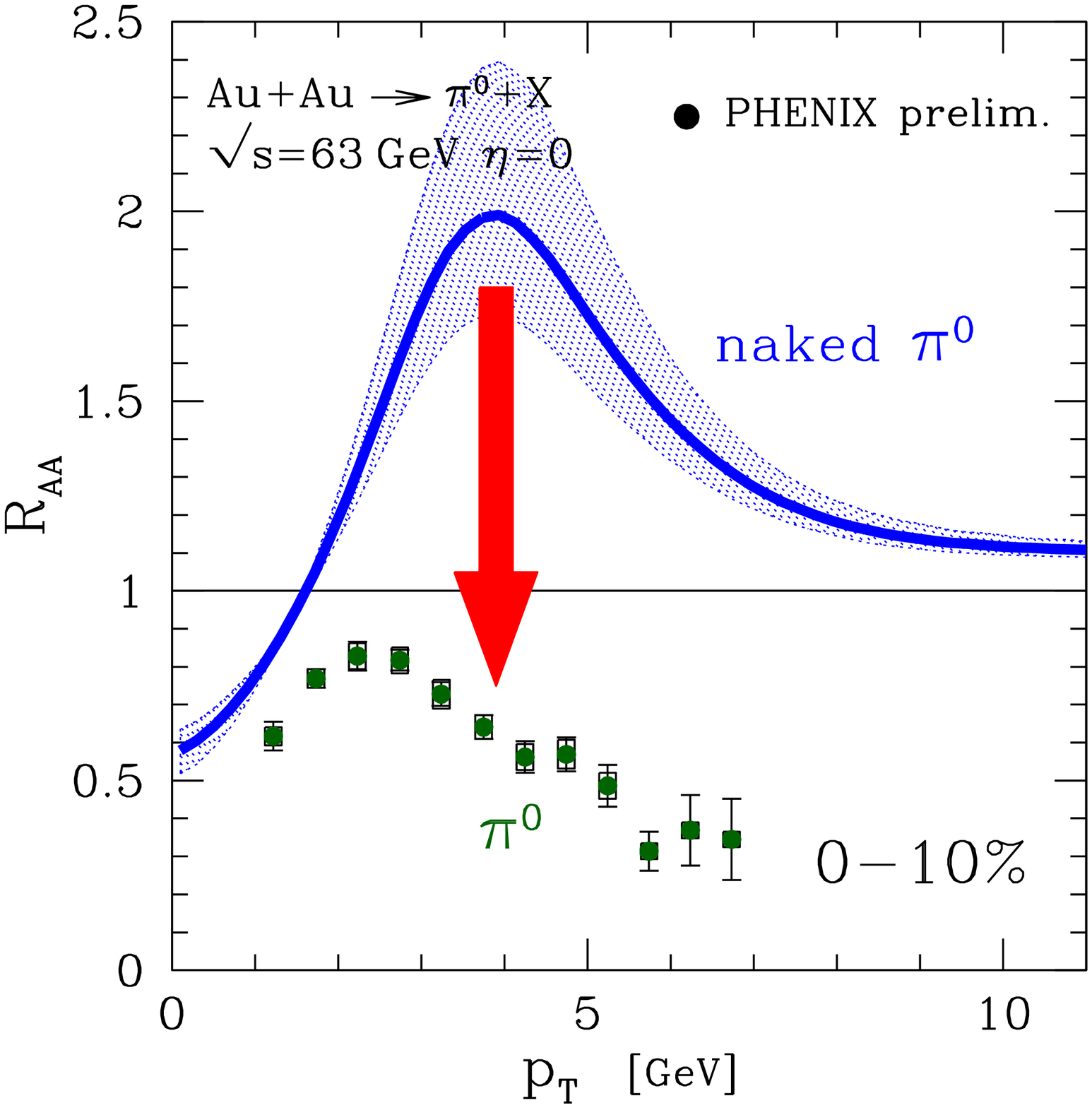,width=0.23\textwidth}
\caption{Naked Cronin effect on $\pi^0$ production in Au+Au collisions
  at $\sqrt s = 62$ GeV compared to preliminary PHENIX experimental data
  from \cite{Bueschingtalk}.
}
\label{fig:PHENIX_AuAu}
\end{figure}

Consider, e.g., the preliminary PHENIX data
on $\pi^0$ production in Au+Au collisions at $\sqrt s = 63$ GeV in
\figref{fig:PHENIX_AuAu}. 
The characteristic bell shape of the Cronin effect at moderate $p_T$
is clearly suppressed. However, the magnitude of the suppression is
not self-evident from experimental data alone: a computation is needed
of the unsuppressed spectrum. The comparison of experimental data and
the GE model computation shows a quite large jet
quenching. Going from central to peripheral collisions, 
the suppressed experimental Cronin peak and the unsuppressed
theoretical curve approach each other. As we can expect this trend to
continue in more and more peripheral events, a natural question
arises: at what centrality does the quenching starts? This is clearly 
a very important question, whose answer will help to pinpoint the
conditions for QGP formation. 

\begin{figure*}[tb]
\epsfig{figure=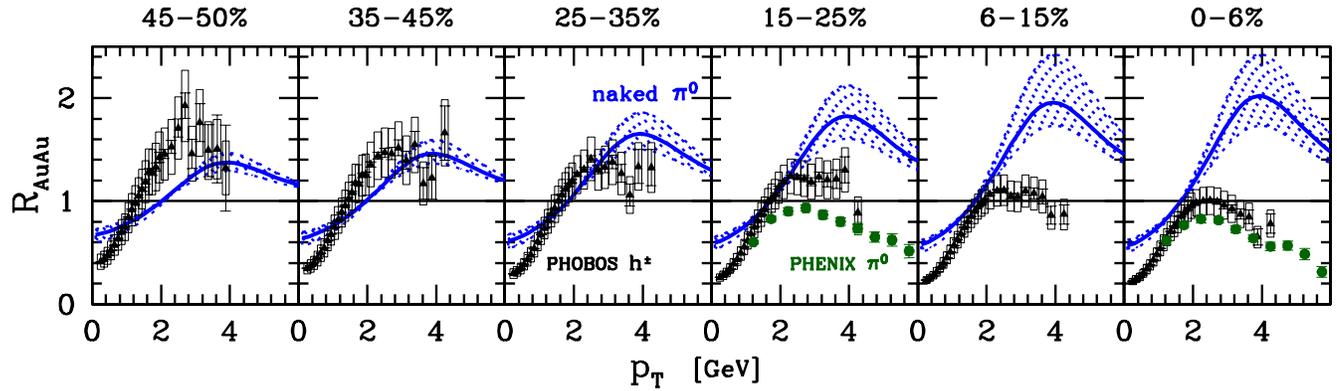,width=\textwidth}
\caption{Naked Cronin effect on $\pi^0$ production in Au+Au collisions
  at $\sqrt s = 62$ GeV compared to experimental
  data for $h^\pm$ from PHOBOS (black triangles) \cite{Phobos63} and
  $\pi^0$ from PHENIX (green disks) \cite{Bueschingtalk}.}
\label{fig:PHOBOS_AuAu}
\end{figure*}

As PHENIX data on $\pi^0$ in more peripheral bins are not yet
available, we may turn to PHOBOS data on charged hadron ($h^\pm$)
production. Here we come across the baryon/meson anomaly previously
discussed: the GE model underestimates the Cronin
effect on protons and anti-protons, which give a sizable contribution
to the $h^\pm$ sample. To attempt nonetheless an answer to the
question of when does the medium start to quench hadrons, we 
may compare $h^\pm$ data to the computation for $\pi^0$'s, see
\figref{fig:PHOBOS_AuAu}. The required modification of the shape and
magnitude of theoretical curves can be estimated by comparing the
PHOBOS $h^\pm$ data points with PHENIX $\pi^0$ data point in similar
centrality bins. With the above caveats in mind, we can estimate that
hadron quenching starts at a centrality of around 40\%. 

I would like to stress the importance of the availability of a
theoretical computation of the naked Cronin effect on the example of
the 4 most peripheral bins in \figref{fig:PHOBOS_AuAu} (up to 15-25\%
centrality). 
The data points do not show a medium effect by themselves: they arise
above 1 at $p_T\approx1.5-2$ GeV, display a nice Cronin peak, and seem
to converge to 1 again at higher $p_T$ inside the systematic errors. 
This behaviour would point to an unsuppressed Cronin effect, but the
comparison to the naked Cronin curve (modulo the baryon/meson anomaly) 
shows a sizable medium effect starting at around 40\%, which otherwise
might have passed unnoticed.

Finally, we can address the WA98 data on $\pi^0$ production in Au+Au
collision at $\sqrt{s}=17$ GeV \cite{WA98-PHENIX}. 
Here we have two complications. 
On the experimental side, there is a large uncertainty on the p+p
spectrum to be used as a normalization in \eqeqref{CroninRatio}, due
to different extrapolations of higher energy data, 
see \refref{dEnterria} for more details and references.  
On the theory side, we have an uncertainty on the $p_0(\sqrt s)$ and
$K(\sqrt s)$ parameters due to the extrapolation at this low 
energy of values extracted \cite{AG04} from data at $\sqrt{s}=27$ and
200 GeV. Preliminary computations show a moderate quenching in central
collisions.

\section{Perspectives}
\label{sec:perspectives}

A cross check of WA98 results may be obtained using the minimum bias
data sample of NA60  \cite{NA60} for In+In collisions at $\sqrt s=17$
GeV. The dimuon trigger data, which has a bias toward central
collisions, may also be used.

NA60 can also perform a nice experimental study of the system size
dependence of the Cronin effect on $h^\pm$ production in p+A collisions 
at low $\sqrt s = 17-27$ GeV.
Indeed, the experimental setup has 7 nuclei (ranging from
beryllium to uranium) mounted in the target region and  
sharing the same beam  \cite{NA60}. This allows to measure the 
ratio $R^\sigma$ of differential cross-sections on two different
nuclei because the beam luminosity cancels in the ratio
\begin{align}
  R^\sigma_{pA/pB} =  
    {\frac{1}{A} \frac{d\sigma}{dp_T^2 dy}}
       ^{\hspace{-0.3cm}pA\rightarrow h X} \Bigg/
    {\frac{1}{B} \frac{d\sigma}{dp_T^2 dy}}
       ^{\hspace{-0.3cm}pB\rightarrow h X} \ ,
  \label{HLratio}
\end{align}
where $B$ is the lightest available nucleus, in this case beryllium. 
The $A$ dependence of the ratio is analogous to the centrality
dependence, but eliminates the
large experimental uncertainties due to the determination of 
the centrality and to the normalization of the Cronin ratio.
 Moreover, without
need of centrality cuts the statistics may be sufficient 
to probe the high-$p_T$ region and test the multi parton scattering
mechanism, which predicts a divergence of the Cronin ratio at
$p_T=4$ GeV and 8 GeV in collision of, viz., $\sqrt s = 17$ GeV and
$27$ GeV.
Theoretical computations will need to take into account the bias
toward central collisions induced by the dimuon trigger, which is
the only trigger used in p+A collisions.

The baryon/meson anomaly is a pervasive phenomenon in QCD: it is
not limited to hadronic and nuclear collisions, but present
also in the HERMES measurements 
of hadron production in deep inelastic scattering on nuclei (nDIS)
\cite{HERMES}.
In this case, the ratio of hadron
multiplicities on a nuclear target and on deuterium is measured as a
function of $z=E_h/\nu$, where $E_h$ is the hadron momentum and
$\nu$ is the virtual photon energy.
Experimentally, meson quenching is
observed on the whole $z=0.2-1$ range, while protons are enhanced at
$z\lesssim 0.4$ and suppressed above; both the quenching and
enhancement increase with $A$. At HERMES the typical hadron momentum 
$E_h \approx 2-12$ GeV is comparable to midrapidity hadron
momenta measured at RHIC:
there is a lot to be learned from nDIS about hadronization, both
experimentally and theoretically, which can be applied to RHIC physics!

\section{Conclusions}
\label{sec:conclusions}

The Glauber-Eikonal model is successful in explaining the Cronin effect
on midrapidity meson production 
in p+A collisions on a wide range of energy, $\sqrt{s}=$27-200 GeV. 
When extended to A+A collisions, it gives a reliable baseline
computation of what the Cronin effect should look like in absence of
the medium-induced hadron quenching. 

By comparing the GE computation with experimental data it is possible
to study the onset and magnitude of hadron quenching. As an example, I
studied recent PHENIX and PHOBOS data in Au+Au collisions at $\sqrt s
= 63$ GeV, and found that the produced medium begins to quench hadron
spectra in collisions of around 40\% centrality, with a strength
quickly increasing with centrality. Preliminary analysis of WA98 data
at $\sqrt s=17$ GeV suggests a 
moderate hadron quenching in central collisions. 
A systematic analysis of all the available experimental 
data is under preparation.  

More can be learned about the Cronin effect in low-energy p+A
collisions from the unique multiple target setup of the NA60 
experiment by studying, as proposed in this note, the $A$-dependence of the
ratio \eqref{HLratio} of differential cross-sections 
on a target nucleus $A$ and on beryllium, the lightest available target. 

The HERMES data on nuclear DIS can shed light on the baryon/meson
anomaly and the space-time evolution of the hadronization process in a
cleaner and more controllable setting than nuclear collisions. 
I would like to invite heavy ion theorists and experimentalists to 
carefully read HERMES papers: there is a lot to be learned which is
directly relevant to RHIC physics. \\

{\bf Acknowledgments.}
Discussions with A.~Dumitru, M.~Gyulassy, J.W.~Qiu and E.~Scomparin
are gratefully acknowledged. This work is partially funded by the US
Department of Energy grant DE-FG02-87ER40371.



\begin{thebibliography}{}
%
\bibitem{Cronin}
J.~W.~Cronin \etal,
Phys.\ Rev.\ D {\bf 11} (1975) 3105;
D.~Antreasyan \etal,
Phys.\ Rev.\ D {\bf 19} (1979) 764.

\bibitem{CroninRHIC}
S.~S.~Adler \etal\  [PHENIX],
Phys.\ Rev.\ Lett.\  {\bf 91} (2003) 072303;
%
J.~Adams \etal\ [STAR],
Phys.\ Rev.\ Lett.\  {\bf 91} (2003) 072304;
B.~B.~Back \etal [PHOBOS],
Phys.\ Rev.\ Lett.\  {\bf 91} (2003) 072302;
I.~Arsene \etal\ [BRAHMS],
Phys.\ Rev.\ Lett.\  {\bf 91} (2003) 072305

\bibitem{CroninAA_SPS}
M.~M.~Aggarwal {\it et al.}  [WA98],
Eur.\ Phys.\ J.\ C {\bf 23} (2002) 225

\bibitem{WA98-PHENIX}
K.~Reygers  [WA98 and PHENIX],
nucl-ex/0202018.

\bibitem{enloss}
M.~Gyulassy, I.~Vitev, X.~N.~Wang and B.~W.~Zhang,
arXiv:nucl-th/0302077.

\bibitem{bmanomaly}
M.~A.~C.~Lamont  [STAR Collaboration],
J.\ Phys.\ G {\bf 30} (2004) S963;
S.~S.~Adler {\it et al.}  [PHENIX],
nucl-ex/0410012.


\bibitem{AG04}
A.~Accardi and M.~Gyulassy,
Phys.\ Lett.\ B {\bf 586} (2004) 244
and 
J.\ Phys.\ G {\bf 30} (2004) S969

\bibitem{Phobos63}
B.~B.~Back {\it et al.} [PHOBOS],
nucl-ex/0409001;
G.~Roland [PHOBOS], these proceedings.

\bibitem{Bueschingtalk}
H.~Buesching [PHENIX], these procedings and 
nucl-ex/0410002.

\bibitem{Hwa}
R.~C.~Hwa,
these proceedings [nucl-th/0501054].

\bibitem{dEnterria}
D.~d'Enterria,
Phys.\ Lett.\ B {\bf 596} (2004) 32
and these proceedings.

\bibitem{NA60} 
See talks of R.~Arnaldi, C.~Louren\c co, R.~Shahoyan and H.~Woehri at this
conference.  
\bibitem{HERMES}
A.~Airapetian {\it et al.}  [HERMES],
Phys.\ Lett.\ B {\bf 577} (2003) 37;
G.~Elbakian \etal\ [HERMES], 
Proceedings of ``DIS 2003'', St.Petersburg, April 23-27, 2003; 
V.T.~Kim and L.N.~Lipatov eds., page 597.  

\end{thebibliography}
\end{document}